\documentclass[12pt,preprint]{aastex}
\slugcomment{accepted by ApJ (Letters) 2009 December 14}
\shortauthors{Geballe \& Oka}
\shorttitle{Remarkable Sightlines into the Galactic Center}

\begin{document}

\title{Two New and Remarkable Sightlines into the GC's Molecular Gas}

\author{T. R. Geballe\altaffilmark{1} and T. Oka\altaffilmark{2}}

\altaffiltext{1}{Gemini Observatory, 670 N. A'ohoku Place, Hilo, HI
96720; tgeballe@gemini.edu}

\altaffiltext{2}{Department of Astronomy and Astrophysics,
                 Department of Chemistry, and Enrico Fermi Institute,
                 University of Chicago, Chicago, IL 60637.}

\begin{abstract}

Until now the known sources in the Galactic center with sufficiently 
smooth spectra and of sufficient brightness to be suitable for high 
resolution infrared absorption spectroscopy of interstellar gas occupied 
a narrow range of longitudes, from the central cluster of hot stars to 
approximately 30 pc east of the center. In order to more fully 
characterize the gas within the r~$\sim$~180~pc central molecular zone 
it is necessary to find additional such sources that cover a much wider 
longitudinal range of sightlines. We are in the process of identifying 
luminous dust-embedded objects suitable for spectroscopy within 1.2$^\circ$ 
in longitude and 0.1$^\circ$ in latitude of Sgr~A* using the {\it 
Spitzer} GLIMPSE and the Two Micron All Sky Survey catalogues. Here we 
present spectra of H$_{3}^{+}$ and CO towards two such objects, one 
located 140 pc west of Sgr A*, and the other located on a line of sight 
to the Sgr B molecular cloud complex 85~pc to the east of Sgr~A*. The 
sightline to the west passes through two dense clouds of unusually high 
negative velocities and also appears to sample a portion of the 
expanding molecular ring. The spectra toward Sgr B reveal at least ten 
absorption components covering over 200~km~s$^{-1}$ and by far the 
largest equivalent width ever observed in an interstellar H$_{3}^{+}$ 
line; they appear to provide the first near-infrared view into that 
hotbed of star formation.

\end{abstract}


\vfill\eject

\section{Introduction}

The Galactic center (GC) is a fascinating environment containing a 
multitude of extraordinary phenomena and extraordinary objects, not the 
least of which are three dense clusters of young and hot stars and a 
multi-million solar mass black hole.  Until recently it was thought that 
the interstellar gas within the central few hundred parsecs of the 
Galaxy, usually referred to as the central molecular zone (hereafter 
CMZ) consisted of three major components \citep{mor96,laz98}: ultra high 
temperature X-ray-emitting plasma; ionized gas at T$\sim$10$^{4-6}$~K 
responsible for the well-studied fine structure and radio recombination 
line emission; and cool and dense molecular clouds, which have also been 
observed in considerable detail at radio wavelengths.  However, recent 
infrared spectroscopy of H$_{3}^{+}$ and CO, and in particular of the 
key $R$(3,3)$^{l}$ absorption line from a metastable state of 
H$_{3}^{+}$ \citep{got02,oka05}, has clearly revealed the presence of 
another component, which in terms of density (50--200 cm$^{-3}$) has the 
characteristics of Galactic diffuse cloud material, but which is 
considerably warmer (200--300~K).  At present, this warm dilute 
environment is unique to the GC; it has not been found in 
any other Galactic diffuse clouds surveyed in H$_{3}^{+}$ (Geballe \& 
Oka, unpublished data). It appears to include gas associated with the 
r~$\sim$~180~pc expanding molecular ring \citep[hereafter 
EMR;][]{kai72,sco72}, which has also been characterized as an expanding 
molecular shell \citep{sof95}, located at the outer edge of the CMZ.

Because of the unique properties of H$_{3}^{+}$ \citep[e.g.,][]{geb06}, 
observations of H$_{3}^{+}$, combined with those of CO, are key to 
characterizing the physical conditions in the CMZ and the extent of the 
warm and diffuse component there. However, spectroscopy of H$_{3}^{+}$ 
is difficult because its lines are weak owing to its low abundance. 
Until recently there has been available as probes of the line of sight 
to the GC only a small number of hot stars in the Central 
cluster and in or near the Quintuplet cluster 30~pc east, which are both 
sufficiently bright for high resolution spectroscopy and have smooth 
infrared spectra so that the H$_{3}^{+}$ line profiles are 
uncontaminated by photospheric absorption lines in the background 
source.

Spectra of these already-known sources \citep{oka05,got08} have shown 
that the warm and diffuse component is present on every sightline and 
also have shown that the H$_{3}^{+}$ column lengths are substantial 
fractions of the radius of the CMZ. They thus suggest that the diffuse 
and warm environment in which the H$_{3}^{+}$ is located takes up a 
large fraction of the volume in the central few hundred parsecs.  If 
correct, this would strongly contradict the previous conceptual picture 
of GC gas, e.g., as illustrated in \citet{laz98}, in which a warm and 
diffuse component has not been included at all.

To better evaluate the extent and physical nature of this newly 
discovered environment, sightlines providing a wider coverage of the CMZ 
are needed. It is therefore essential to find additional bright sources 
with featureless or nearly featureless spectra -- either hot stars with 
few emission or absorption lines, or stars encased in dense shells of 
warm dust -- in a more extended region of the GC.

\section{Finding New Sightlines through the CMZ}

The CMZ is filled with bright infrared sources, but everywhere except at 
locations of the three clusters of luminous hot stars (the Central, 
Arches, and Quintuplet clusters) the overwhelming majority of them are 
red giants, whose complex photospheric absorption spectra make them 
unsuitable as probes of the interstellar medium. Until very recently, no 
smooth-spectrum objects in the line of sight to the CMZ but far from 
those clusters were known. We are using the Two Micron All Sky Survey 
(2MASS) Point Source catalogue \citep{skr06} and the {\it Spitzer Space 
Telescope} GLIMPSE catalogue \citep{ram08} to identify bright objects in 
the direction of the CMZ that are likely to have opaque dust shells. A 
simplified description of the technique is that the shorter wavelength 
IR colors are mainly used to weed out foreground (low extinction) 
sources and ``normal" red giants, and the longer wavelength IR colors 
are mainly used to identify emission from warm dust.  However, the 
situation is far from straightforward, because the effects on 
1--8~$\mu$m photometry of high extinction and low temperature cannot be 
easily separated. Our success rate, although much higher than a random 
sampling, is currently only $\sim$15\%.

Thus a check of each candidate is necessary before proceeding to the 
time-consuming high-resolution spectroscopy. This second step is 
performed by obtaining quick medium-resolution $K$-band spectra, in 
particular covering the first overtone CO bands at 2.3--2.5~$\mu$m, to 
determine if the candidates do indeed have smooth spectra or are cool 
red giants suffering high extinction.

Our requirement for high resolution spectroscopy of the key lines of 
H$_{3}^{+}$, which mostly lie in the 3.5--3.7~$\mu$m region, is that the 
sources have Infrared Array Camera (IRAC) band 1 (3.6~$\mu$m) magnitudes 
brighter than 8.  Roughly 2,000 GLIMPSE sources with 
$-$1.2$^\circ$~$<$~$l$~$<$~+1.2$^\circ$ and 
$-$0.1$^\circ$~$<$~$b$~$<$~+0.1$^\circ$ (here $l$ and $b$ are offsets in 
Galactic longitude and latitude from Sgr~A*, assumed to be at a distance 
of 8.0~kpc) satisfy that criterion. Most of them have 2MASS 
counterparts. Based on 2MASS $J-K$, 2MASS/Spitzer $K$$-$IRAC(1) and 
Spitzer IRAC(1)$-$IRAC(4) (8.0~$\mu$m) colors, we have compiled a list 
of $\sim$250 candidate dusty sources that to our knowledge had not 
previously been observed spectroscopically. $K$-band spectra of 
approximately 75 of them now have been obtained. Of those, ten, whose 
locations are shown in Figure~1, have been found to be suitable for high 
resolution spectroscopy of interstellar gas lines. We have no additional 
information concerning the natures of these ten sources. They are likely 
to contain either young stellar objects or luminous evolved stars.

At the stage that the $K$-band spectroscopy has revealed suitable 
sources, the locations of those sources along the line of sight are 
unknown.  Although we attempt to select for high interstellar 
extinction, it is quite possible that some of the sources are situated 
in front of the GC. High resolution spectroscopy of CO first overtone 
lines originating in low $J$ levels of the ground vibrational state can 
help to locate the sources on the line of sight. Previous observations 
by \citet{oka05} have demonstrated that the spectra of objects in the GC 
show narrow absorption components of H$_{3}^{+}$ and CO arising in 
foreground spiral arms. The presence or absence of absorption components 
at the characteristic velocities of these foreground arms can provide 
useful constraints. However, the clouds along the intervening spiral 
arms may not be continuous, but instead clumpy on small scales. Thus the 
lack of an absorption at a velocity characteristic of a spiral arm does 
not necessarily prove that the object is located in front of that arm.

Despite the low efficiency and the possibilities of confusion about 
location on the line of sight, the technique already shows great promise 
of providing a more extensive and more unifom sampling of the molecular 
gas in the CMZ than previously available. In particular, the sightlines 
toward two of newly found objects, 2MASS J174332173$-$2951430 and 2MASS 
J17470898$-$2829561 (hereafter 2M1743 and 2M1747, respectively), contain 
remarkable collections of interstellar clouds absorbing in lines of CO 
and H$_{3}^{+}$. In the following sections we describe the exploratory 
spectra we have obtained of them.

\section{Medium Resolution Spectra}

Medium-resolution 1.4--2.5~$\mu$m spectra of 2M1743 ($K$~=~6.5) and 
2M1747 ($K$~=~10.4) were obtained at the United Kingdom Infrared 
Telescope (UKIRT) on Mauna Kea on 2008 July 28 and August 15, 
respectively, using the facility imager/spectrograph UIST, whose 
0.2\arcsec\ wide slit provided a resolving power of 1000. On both 
nights HR~6409 (F6~IV) was observed at roughly the same air mass as the 
2MASS objects for the purposes of flux calibration and removal of 
telluric lines. Total integration times on the 2MASS objects were 80 and 
360 seconds, respectively. Observations were made in 
stare/nod-along-slit mode. Data reduction was standard for near-infrared 
spectroscopy of point sources. Wavelength calibration was obtained from 
telluric absorption lines observed in HR~6409 and is accurate to better 
than 0.0005~$\mu$m. The 2.166~$\mu$m Br~$\gamma$ absorption line in 
HR~6409 was removed by interpolation prior to ratioing.

The 2.0--2.4~$\mu$m portions of the spectra of the two objects are shown 
in Fig.~2. 2M1743 has a smooth and steeply rising spectrum, consistent 
with that of a dust-embedded star. The spectrum of 2M1747, which rises 
even more steeply, is also indicative of warm dust. However, while the 
spectrum of 2M1743 appears featureless at this resolution, that of 
2M1747 shows several significant absorptions. These include the 2--0 and 
3--1 band heads of CO, perhaps originating in the veiled photosphere of 
a cool and luminous star or in a dense and high-temperature 
circumstellar shell or disk of a young stellar object. In addition, 
significant absorption is seen near the wavelength of the 2--0 CO band 
center (2.347~$\mu$m), suggesting the presence of an unusually large 
column density of lower temperature (interstellar) CO. Finally, a broad 
absorption band, centered at approximately 2.265~$\mu$m, is present. It 
has a full width at zero intensity of $\sim$~0.02~$\mu$m. We are unable 
to identify this feature. Its wavelength range encompasses that of the 
triplet of neutral calcium lines seen in late-type stars \citep{kle86}; 
however, the feature is too broad and too strong relative to CO for that 
identification to be viable. It is possible that the absorption is 
produced in frozen grain mantles within molecular clouds along the line 
of sight. An absorption at 2.27~$\mu$m with a similar profile, possibly 
due to solid methanol, has been observed in some solar system objects 
\citep{cru98}.

\section{High Resolution Spectra of H$_{3}^{+}$ and CO}

High resolution spectra of both objects at the $R$(1,1)$^{l}$ transition 
of H$_{3}^{+}$ near 3.715~$\mu$m and covering a small portion of the 
2--0 band of CO near 2.342~$\mu$m were obtained at the Gemini South 
telescope on Cerro Pachon in Chile on 2009 July 6. The observations used 
the echelle spectrograph, Phoenix, whose 0.34\arcsec\ wide slit provides 
a resolving power of 50,000. In one setting the spectral coverage 
corresponds to $\Delta$$\lambda$/$\lambda$~=~0.0045 on the instrument's 
detector array. For the CO spectra the echelle was centered at 
2.342~$\mu$m, thereby covering the five lowest lying $R$ branch 
transitions of the 2--0 band, i.e., $R$(0)--$R$(4). The separation of 
adjacent 2--0 rovibrational CO lines corresponds to a velocity range of 
260~km~s${-1}$; thus if the absorption profile is broad the baseline for 
defining the continuum level between CO lines is restricted.  The other 
setting was centered on the wavelength of the H$_{3}^{+}$ line, whose 
lower level is the ground state. Data reduction was similar to that 
described earlier, with HR~6070 (A0V) and HR~7254 (A2V) serving as 
standards for both wavelength intervals. Wavelength calibrations used 
telluric absorption lines, and the resultant velocity scales in Figures 
3 and 4 are accurate to 2~km~s$^{-1}$.

\subsection{2MASS J17432173$-$2951430}

Profiles of the H$_{3}^{+}$ line and the CO $R$(0)--$R$(3) lines 
observed toward 2M1743 are shown in Figure~3. Absorption components of CO 
are present at LSR velocities of $-$60, $-$172, and $-$200 km~s$^{-1}$. 
The $-$172 km~s$^{-1}$ absorption profile is slightly asymmetric, 
indicating the presence of a second and weaker absorption red-shifted by 
a few km~s$^{-1}$. The H$_{3}^{+}$ $R$(1,1)$^{l}$ spectrum also contains 
prominent absorption components, including the same three seen in CO, 
and a red-shifted shoulder on the $-$172 km~s$^{-1}$ absorption that is 
relatively stronger than in CO. A fourth prominent absorption in the 
H$_{3}^{+}$ spectrum, which is not present in CO, is an apparent 
velocity doublet at 0 and +8 km~s$^{-1}$. Finally, broad but weaker 
H$_{3}^{+}$ absorptions, which also have no counterparts in CO, are 
centered near $-$27 and $-$75~km~s$^{-1}$.

The CO absorptions observed at $-$60, $-$172, and $-$200 km~s$^{-1}$ are 
likely to be formed in dense clouds. Only the first four rotational 
levels are significantly populated. The overall CO excitation 
temperature is roughly 10~K, but it is quite possible that the kinetic 
temperature is higher and that the level populations are sub-thermal. A 
more thorough analysis will be provided in a subsequent paper. The 
component at -60~km~s$^{-1}$ possibly arises in the 3~kpc arm. On 
sightlines much closer to the center the absorption ascribed to that 
spiral arm occurs near $-$52~km~s$^{-1}$ \citep{oka05}.  The other two 
CO components, at much higher velocity, do not correspond to foreground 
spiral arms. Because of their high velocities it is likely that these 
features arise close to the GC. CO $J$=1--0 spectra obtained by 
\citet{oka98} approximately along this sightline 
($l_{II}$~=~358.954$^\circ$, $b_{II}$~=~$-$0.066$^\circ$) have their 
strongest emission components at those two high negative velocities.

Because no infrared CO absorption is present at the velocities of the 
H$_{3}^{+}$ absorptions near 0, +8, and $-$75~km~s$^{-1}$, the clouds 
producing them must be diffuse. Only weak CO $J$=1--0 line emission is 
present at those velocities. 2M1743 is located within a few parsecs of 
the galactic plane and is approximately 140~pc west of the center (see 
Figure~1). If it is located somewhat behind the center, its sightline 
would cross the EMR where its gas is moving nearly in the plane of the 
sky (with very little Doppler shift). It is thus logical to associate 
the low velocity doublet with the EMR and to place 2M1743 somewhat 
behind the EMR. Previous observations have demonstrated that the EMR 
contains diffuse gas \citep{oka05,got08}. If the identification is 
correct, it is evidence that the diffuse nature of the EMR's gas is 
widespread, and is not limited to the sightlines close to the longitudes 
of the Central and Quintuplet clusters.

We have no specific identification for the H$_{3}^{+}$ features at 
$-$27~km~s$^{-1}$ and $-$75~km~s$^{-1}$. However, previously observed 
GC slightlines \citep{oka05,got08} showed a trough of 
absorption by diffuse gas from 0 to $-$100~km~s$^{-1}$, indicating that 
a significant fraction of the volume of the CMZ contains diffuse gas. If 
so, then the presence of additional H$_{3}^{+}$ absorption components 
in that velocity range would not be surprising.

\subsection{2MASS J17470898$-$2829561}

Velocity profiles of the H$_{3}^{+}$ line and the CO $R$(0)--$R$(3) 
lines toward 2M1747 are displayed in Figure~4. Both molecules absorb 
continuously over wide velocity ranges. Absorption by CO extends without 
interuption from $-$100 to +100~km~s$^{-1}$, and the absorption by 
H$_{3}^{+}$ extends even further without a break, from $-$130 to 
+100~km~s$^{-1}$. About a dozen discrete velocity components can be 
seen in both the H$_{3}^{+}$ and CO line profiles. Several, but not all 
of the components of the two molecules coincide, and thus the sightline 
appears to contain a combination of diffuse and dense clouds, but at 
present it is not possible to untangle the two contributions. The only 
clear indication of gas with a low excitation temperature similar to 
that seen in the CO toward 2M1743 is at $-$43~km~s$^{-1}$, where the 
strongest absorption occurs in the $J$~=~0, 1, and 2 levels and where 
the CO absorption depth noticeably decreases with increasing lower state 
energy. This absorption component may be a continuation of absorption by 
molecular gas in the 3~kpc arm, as discussed previously for 2M1743.

At $l_{II}$~=~0.548$^\circ$, $b_{II}$~=~$-$0.060$^\circ$, 2M1747 is 
located approximately 85~pc east of the Galactic center on the line of 
sight to the Sgr B giant molecular cloud complex, and is almost directly 
between Sgr~B1 and Sgr~B2 (Figure~1). The CO $J$=1--0 spectra of 
\citet{oka98} at this location shows a strong and complex emission 
profile at positive velocities, not very different from those seen in 
the infrared CO and H$_{3}^{+}$ lines, but very little emission at 
negative velocities where both the infrared CO and H$_{3}^{+}$ 
absorptions also are strong.

Given the unprecedented large widths of the H$_{3}^{+}$ $R$(1,1)$^{l}$ 
and interstellar CO lines (that of the H$_{3}^{+}$ line is roughly twice 
that previously reported for any other GC sightline), it seems beyond 
doubt that 2M1747 lies within Sgr~B. To our knowledge the absorption 
spectra in Figure~4 are the first near-infrared views into that complex 
and turbulent star-forming region.

\section{Conclusion}

The spectra presented here represent the beginning of a new phase of 
exploration of the CMZ using absorption spectroscopy of H$_{3}^{+}$ and 
CO along new sightlines, which has already yielded striking results. 
More detailed understanding of the gas on these two new sightlines, as 
well as on others that have been or are likely to be found, will require 
spectroscopy of additional transitions of H$_{3}^{+}$, in particular of 
the $R$(2,2)$^{l}$ and $R$(3,3)$^{l}$ lines, arising from higher energy 
levels than the $R$(1,1)$^{l}$ line, and detailed comparison with 
infrared and millimeter spectra of CO and perhaps other molecular 
species.

\begin{acknowledgements}

Some of the data presented here were obtained at the Gemini Observatory, 
which is operated by the Association of Universities for Research in 
Astronomy, Inc., under a cooperative agreement with the NSF on behalf of 
the Gemini partnership: the National Science Foundation (USA), 
the Science and Technology Facilities Council (UK), the 
National Research Council (Canada), CONICYT (Chile), the Australian 
Research Council (Australia), Minist\'erio da Ci\'encia e Tecnologia 
(Brazil) and SECYT (Argentina). The remaining data were obtained at 
UKIRT, which is operated by the Joint Astronomy Centre on behalf of the 
U.K. Science and Technology Facilities Council. The authors thank the 
staffs of both institutions for their support of this work. We also 
thank Tomoharu Oka for providing us with his CO $J$=1--0 spectra. TO is 
supported by NSF grant AST-0849577.

\end{acknowledgements}

\clearpage

\begin{figure} 
\epsscale{0.8}
\plotone{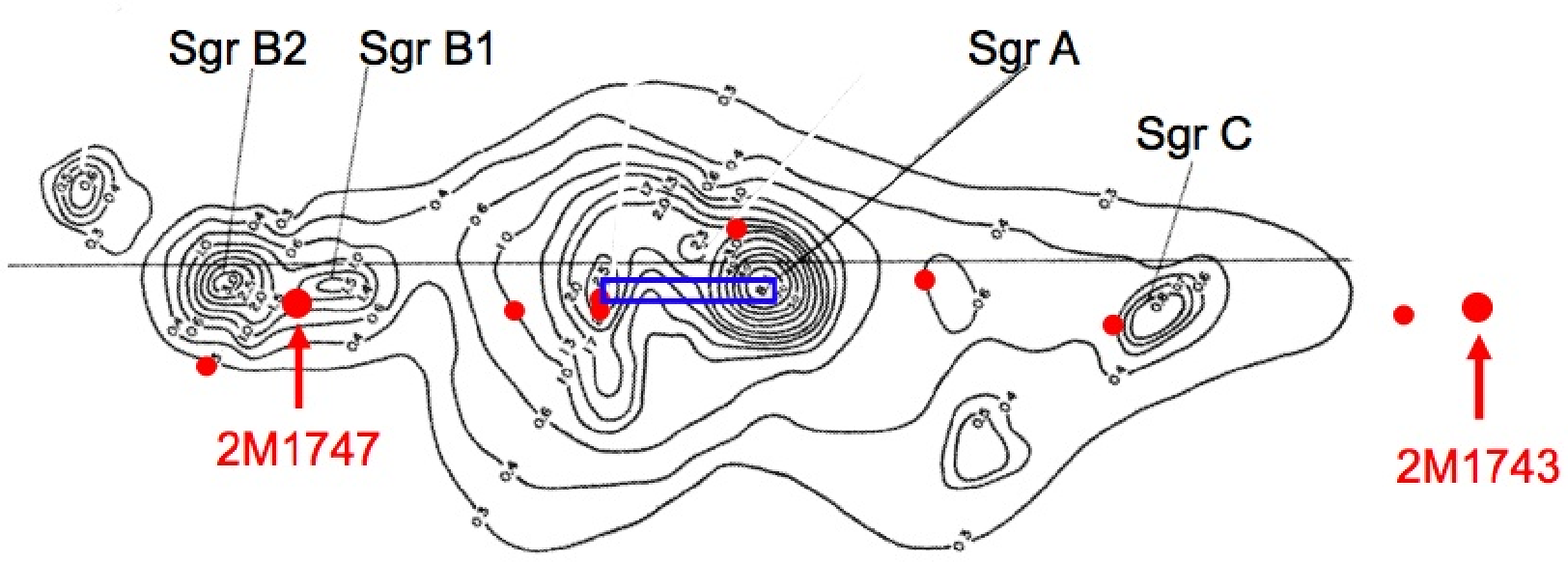} 
\label{radio} 
\caption{A portion of a radio contour map of the GC at 8.0 GHz 
\citep{dow66}, with principal radio sources labelled. Locations of 
2MASS~J17432173$-$2951430 and 2MASS~J17470898$-$2951430 are indicated by 
large red dots and arrows; locations of other newly discovered smooth 
spectrum 2MASS sources are indicated by small dots. Locations of sources 
previously observed in H$_{3}^{+}$ lines fall within the narrow blue 
rectangle. The horizontal line denotes the Galactic plane 
($b_{II}$~=~0$^\circ$).}
\end{figure}
\clearpage

\begin{figure} 
\epsscale{0.8} 
\plotone{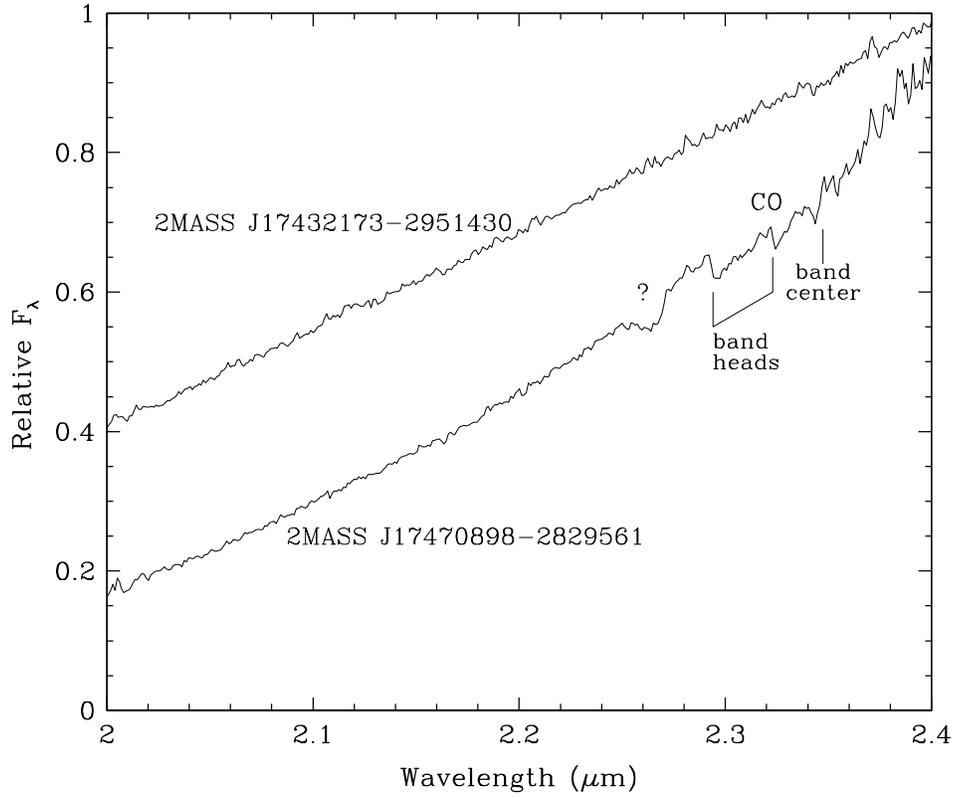} 
\label{uist}
\caption{Medium resolution 2.0--2.4~$\mu$m spectra of two sources 
identified as likely dust-embedded stars located close to the GC. 
Locations of spectral features are indicated and identified if 
known.}
\end{figure}
\clearpage

\begin{figure} 
\plotone{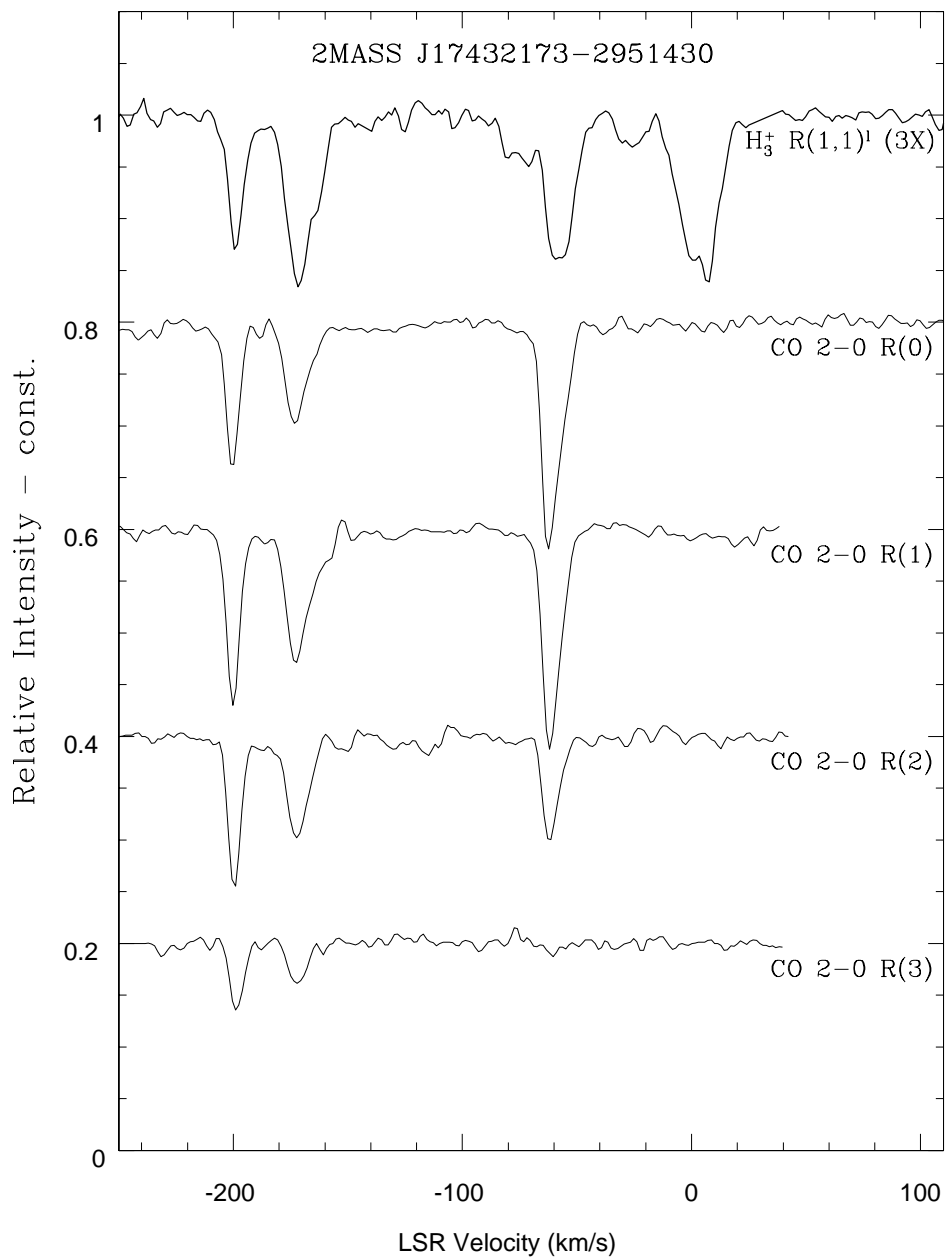}
\label{2M1743} 
\caption{Spectra of the $R$(1,1)$^{l}$ line of H$_{3}^{+}$ and the four 
lowest lying transitions of the 2$-$0 $R$ branch of CO in 
2MASS~J17432173$-$2951430. Spectra are offset vertically. CO spectra are 
to the same scale; an opaque CO line would have depth unity. The 
H$_{3}^{+}$ spectrum is magnified by a factor of 3. Noise can be 
judged by point-to-point fluctuations in flat regions of the spectra.}
\end{figure}
\clearpage

\begin{figure}
\plotone{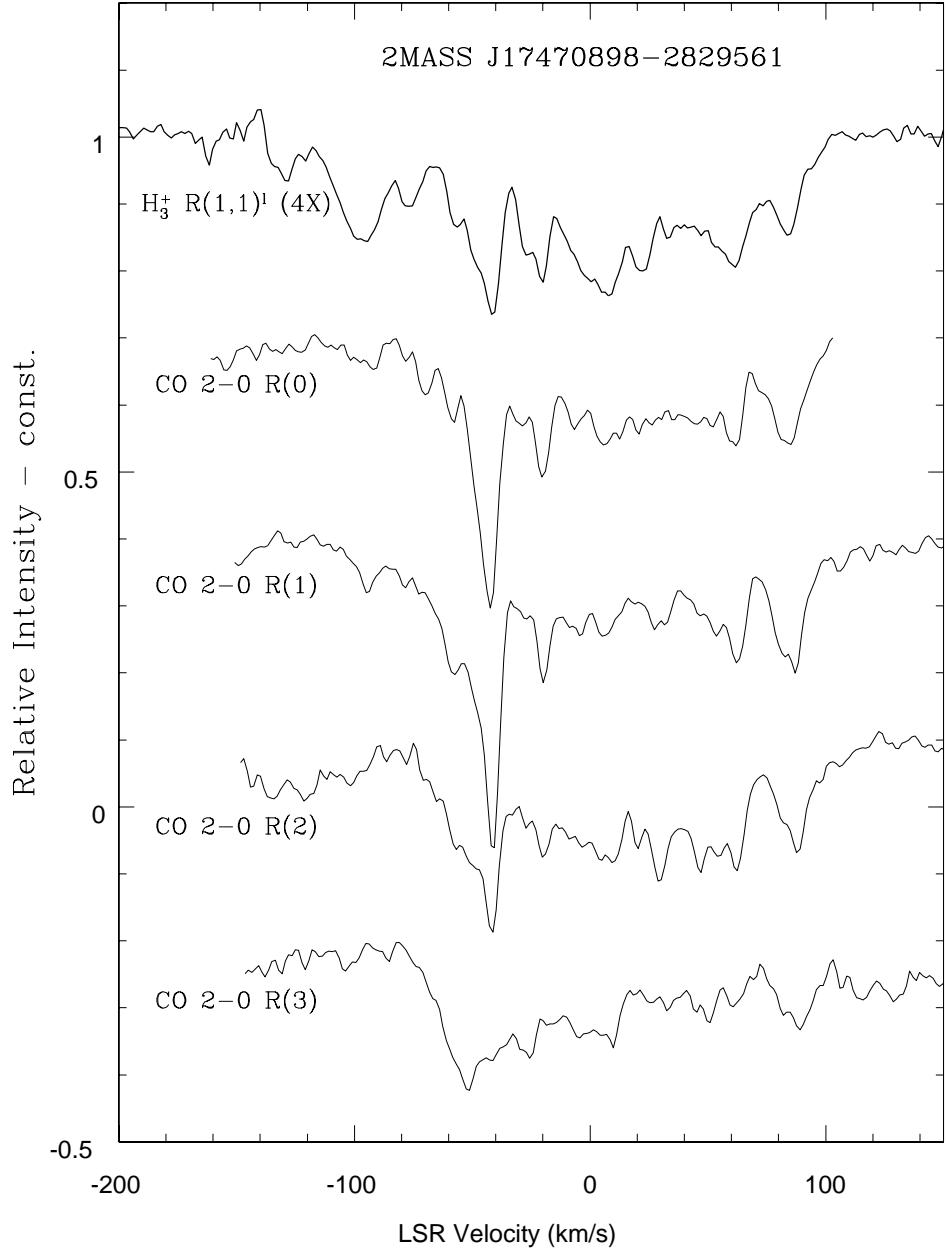}
\label{2M1747}
\caption{Spectra of the $R$(1,1)$^{l}$ line of H$_{3}^{+}$ and the four 
lowest lying transitions of the 2$-$0 $R$ branch of CO in 
2MASS~J17470898$-$2951430.  CO spectra are to the same scale; an opaque 
CO line would have depth unity. The H$_{3}^{+}$ spectrum is magnified by 
a factor of 4.}
\end{figure} 
\clearpage

\end{document}